\documentclass[preprint,12pt]{elsarticle}
\usepackage[utf8,latin1]{inputenc}
\usepackage{color}
\usepackage{amsmath,amssymb,amsfonts,amsthm}

\newcommand{\vw}{{\mathbf{w}}}

\newcommand{\vf}{{\mathbf{f}}}

\newcommand{\cT}{{\mathcal T}}

\newcommand{\beq}{\begin{equation}}
\newcommand{\eeq}{\end{equation}}
\newcommand{\be}{\begin{equation}}
\newcommand{\ee}{\end{equation}}
\newcommand{\bes}{\begin{equation*}}
\newcommand{\ees}{\end{equation*}}
\newcommand{\bea}{\begin{eqnarray}}
\newcommand{\eea}{\end{eqnarray}}
\newcommand{\beas}{\begin{eqnarray*}}
\newcommand{\eeas}{\end{eqnarray*}}
\newcommand{\beqn}{\begin{eqnarray}}
\newcommand{\eeqn}{\end{eqnarray}}
\newcommand{\beqns}{\begin{eqnarray*}}
\newcommand{\eeqns}{\end{eqnarray*}}

\def\fr#1#2{\frac{#1}{#2}}
 \usepackage{graphics}
\usepackage{amssymb}
\newcommand{\bv}{{\bf v}}
\newcommand{\bx}{{\bf x}}
\newcommand{\eps}{\varepsilon}
\newcommand{\cQ}{{\mathcal Q}}
\journal{European Journal of Mechanics - B/Fluids}

\begin{document}
\begin{frontmatter}



\title{Fully nonlinear weakly dispersive modelling of wave transformation, breaking and runup}


\author[L1]{Bonneton, P.}
\ead{p.bonneton@epoc.u-bordeaux1.fr}
\author[L2]{Barthelemy, E.}
\author[L3]{Carter, J.D.}
\author[L4]{Chazel, F.}
\author[L5]{Cienfuegos, R.}
\author[L6]{Lannes, D.}
\author[L7]{Marche, F.}
\author[L1]{Tissier, M.}
\address[L1]{Universit\'e Bordeaux 1; CNRS; UMR 5805-EPOC, Talence, F-33405 France }
\address[L2]{LEGI, INP,
BP53 38041 Grenoble Cedex 9, France.}
\address[L3]{Mathematics Department, Seattle University, Seattle, WA 98122, USA }
\address[L4]{Universit\'e de Toulouse, UPS/INSA, IMT, CNRS UMR 5219, F-31077 Toulouse, France}
%\ead{florent.chazel@math.univ-toulouse.fr}
\address[L5]{Departamento de Ingenier\'ia Hidr\'aulica y Ambiental,
Pontificia Universidad Catolica de Chile,
Vicu\~{n}a Mackenna 4860, casilla 306, correo 221, Santiago de Chile.}
\address[L6]{DMA/CNRS UMR 8553, Ecole Normale Sup\'erieure, 45 rue d'Ulm, 75005 Paris, France}
\address[L7]{I3M, Universit\'e Montpellier 2, CC 051, F-34000 Montpellier, France}
%\ead{fabien.marche@math.univ-montp2.fr}

\begin{abstract}
To describe the strongly nonlinear dynamics of waves propagating in the final stages of shoaling and in the surf and swash zones, fully nonlinear models are required. The ability of the Serre or Green Naghdi (S-GN) equations to reproduce this nonlinear processes is reviewed. Two high-order methods for solving S-GN equations, based on Finite Volume approaches, are presented. The first one is based on a quasi-conservative form of the S-GN equations, and the second on a hybrid Finite Volume/Finite Difference method. We show the ability of these two approaches to accurately simulate nonlinear shoaling, breaking and runup processes.
\end{abstract}

\end{frontmatter}

%%
%% Line numbers can be started here if you want.
%%
% \linenumbers


%
%%%%%%%%%%%%%%%%%%%%%%%%%%%%%%%%%%%%%%%%%%%%%%%%%%%%%%%%%%%%%%%%%%%%%%%%%%%%%%%%%%%%%%%%%%%%%%%%%%%%%%%%%%%%%%%%%%%%%%%%%%%
%
%\tableofcontents
\section{Introduction}
\label{intro}

Wave propagation in shallow water, and associated processes such as wave-breaking and run-up, play an important role in the nearshore dynamics. The classical and successful method of describing slowly evolving wave-induced circulation in the nearshore is based on the phase-averaged approach, in which the depth-integrated mass and momentum equations are time-averaged over a wave period (see Phillips \cite{phillips77}). However, very important unsteady processes, such as wave run-up in the swash zone, coastal flooding during storm, tsunami and tidal bore propagation, require phase-resolving models. For coastal applications, these models are based on nonlinear shallow water equations (NSWE) and Boussinesq-type equations (see Brocchini and Dodd \cite{broc2008}). NSWE give a good description of the nonlinear non-dispersive transformation of broken-waves, represented as shocks, in the inner surf and swash zones. However, due to the absence of frequency dispersion, the NSWE can not be applied to wave propagation before breaking. On the other hand, Boussinesq equations incorporate frequency dispersion and can be applied to wave shoaling, but contrary to NSWE they do not implicitly take into account wave breaking. Since the 1990's, significant efforts have been devoted to extend the validity range of Boussinesq equations, by developing wave breaking parametrizations and by improving dispersive properties of these equations. Most of Boussinesq models used for nearshore applications are based on  classical assumptions of weak nonlinearity, $\epsilon =a/h_0 \ll 1$ ($a$ of the order of free surface amplitude, $h_0$ the characteristic water depth)  and balance 
between dispersion and nonlinearity: $\epsilon=O(\mu)\ll 1$, where $\mu=(h_0/L)^2$ ( $L$ the characteristic horizontal scale). However, these assumptions may severely restrict applicability to real nearshore applications. Indeed, in the final stages of shoaling or in the surf zone, the wave dynamics is strongly nonlinear: $\epsilon=O(1)$. For instance, $\epsilon$ is close to $0.4$  just before breaking  and can be larger than $1$ in the swash zone. 

In 1953, a breakthrough treating nonlinearity was made by Serre (see Barth\'elemy \cite{bar2004} for a review). He derived 1D fully nonlinear ($\epsilon=O(1)$) weakly dispersive equations for horizontal bottom. Green and Naghdi \cite{gre1976} derived 2D fully nonlinear  weakly dispersive 
equations for uneven bottom which represent a two-dimensional extension of Serre equations. 
Except for being formulated in terms of the velocity vector at an arbitrary $z$ level, 
the equations of Wei et al. \cite{wei1995} are basically equivalent to the 2D Serre 
or Green-Naghdi equations.  It is now recognized that the Serre or Green Naghdi (S-GN) equations represent the relevant system to model highly nonlinear weakly dispersive waves propagating in shallow water (see Lannes and Bonneton \cite{lann_POF2009}). However, much remains to be done for a proper representation of wave breaking, and for an accurate modeling of moving shorelines over strongly varying topographies. 

Due to facility of implementation, most of the Boussinesq-type models use finite difference schemes to discretize the equations (e.g. Abbott et al. \cite{abbott1984} , Wei et al. \cite{wei1995}). 
Finite volume methods, which are derived  on the basis of the integral form of the conservation law, have many advantages. They are conservative and easily formulated to allow for unstructured meshes. 
Although the finite volume method has been widely and successfully used to solve the strictly hyperbolic NSWE (e.g. Leveque \cite{LeVeque2002}, Brocchini and Dodd \cite{broc2008}, Marche et al. \cite{marc2007}), its application to Boussinesq-type equations has only been  reported recently (Bradford and Sanders \cite{bradford2002}, Stansby \cite{stansby2003}). This is related to the fact that finite volume methods essentially aim at a good representation of advection, while methods used for Boussinesq-type equations must also deal with third order derivatives responsible for dispersive effects.  To overcome this problem, hybrid approaches, coupling finite volume and the finite difference methods have been recently proposed (Soares Frazao and  Zech \cite{soares2002}, Bernetti et al. \cite{bernetti2003}, Erduran et al. \cite{erduran2005}). The hyperbolic terms are treated  using shock-capturing methods while the dispersive terms are discretized using the finite difference formulation. Until now, in coastal applications finite volume and hybrid approaches have only been applied for weakly nonlinear forms of Boussinesq-type equations. However, in the final stages of shoaling and in the surf and swash zones, the effects of nonlinearity are too large to be treated as a small perturbation. In the context of the IDAO ocean research programme we have investigated, the last few years, applications of finite volume and hybrid methods to the fully nonlinear weakly dispersive S-GN equations. In this paper, we present a synthesis of this work, emphasizing  the ability of S-GN models to deal with  wave transformation in the surf and swash zones. 

%
%%%%%%%%%%%%%%%%%%%%%%%%%%%%%%%%%%%%%%%%%%%%%%%%%%%%%%%%%%%%%%%%%%%%%%%%%%%%%%%%%%%%%%%%%%%%%%%%%%%%%%%%%%%%%%%%%%%%%%%%%%%
%
\section{Theoretical background}
\label{theory}
%
%%%
%
According to \cite{alvarezlannes} and \cite{lann_POF2009}, the 2D S-GN equations can be written under the following non-dimensionalized form
\begin{eqnarray}
\nonumber
\zeta_t+\nabla\cdot(h\bv)&=&0\\
	\label{eqGN}
\bv_t+\eps (\bv\cdot\nabla)\bv+\nabla\zeta &=& - \mu {\bf\cal D} \ ,
\end{eqnarray}
where $\zeta(\bx,t)$ is the surface elevation, $h(\bx,t)=1+\epsilon \zeta - b$ the water depth, $b(\bx)$ the variation of the bottom topography and $\bv(\bx,t)=(u,v)$ the depth averaged velocity. $\cal D$ characterizes non-hydrostatic and dispersive effects and writes 
\begin{equation}\label{eqD}
 {\bf\cal D}={\mathcal T}[h,b]\bv_t+  \eps \left( -\frac{1}{3h}\nabla\left(h^3((\bv\cdot \nabla) (\nabla\cdot \bv)-(\nabla\cdot \bv)^2)\right)+\cQ[h,b](\bv) \right) 
\end{equation}
where the linear operator 
${\mathcal T}[h,b]$ is defined as 
\begin{equation}\label{eqT0}
	{\mathcal T}[h,b] W=
	-\frac{1}{3h}\nabla(h^3\nabla\cdot W)+\frac{1}{2h}\big[\nabla(h^2\nabla b \cdot W)-h^2\nabla b \nabla\cdot W\big]+ \nabla b\nabla b\cdot W \,
\end{equation}
and the purely topographical term 
$\cQ[h,b](\bv)$  is given by
\begin{eqnarray}
	\nonumber
	\cQ[h,b](\bv)
	&=&\frac{1}{2h}\big[\nabla\big(h^2(\bv\cdot\nabla)^2b\big)
	-h^2\big((\bv\cdot \nabla) (\nabla\cdot \bv)
	-(\nabla\cdot \bv)^2\big)\nabla b\big]\\
	\label{eqcQ}
	& &+\big((\bv\cdot\nabla)^2b\big)\nabla b.
\end{eqnarray}

The range of validity of this set of equations can be easily extended to wave
propagation problems in deeper waters using the dispersion correction technique discussed in 
references \cite{wit1984}, \cite{mad1991}, \cite{cien_bb_ijnmf_1_2006} and \cite{bclmt2009}. The frequency dispersion properties are improved by  applying the operator $I+\mu(\alpha-1){\mathcal T}[h,b]$ to the momentum equation (\ref{eqGN}) and neglecting $O(\mu^2)$ terms, which makes the term,
\begin{equation}
(\alpha-1) \mu S = - (\alpha-1)\mu{\mathcal T}[h,b](\bv_t+\eps (\bv\cdot\nabla)\bv+\nabla\zeta) \ ,
\label{eq_disp}
\end{equation}
appears on the right-hand side of the momentum equation (\ref{eqGN}).\\
 The coefficient $\alpha$ is an adjustable parameter that must be tuned in order to minimize the phase and group velocity errors in comparison with the linear Stokes theory. This yields the optimal value $\alpha=1.159$ \footnote{the parameter used in reference \cite{cien_bb_ijnmf_2_2007} is given by $\alpha'=(\alpha-1)/3$, with an optimal value $\alpha' =0.053$}. Inspired by Nwogu \cite{Nwogu}, it is also possible to choose another dependent variable (such as the velocity at a certain depth) rather than the mean velocity, as in \cite{wei1995,CLM}. This allows for more freedom to match the Stokes linear dispersion and/or the exact linear shoaling coefficient. The price to pay is that the mass conservation equation is not exact anymore, but of order $O(\mu^2)$.

It is known that the S-GN equations are mathematically well-posed in the sense that they admit solutions over the relevant time scale for any initial data reasonably smooth (see \cite{alvarezlannesB} for the general case, and simpler proofs for one dimensional surfaces \cite{Li,Israwi}). Moreover, the solution of the S-GN equations provides a good approximation of the solution of the full water waves equations (see \cite{alvarezlannes} for the general case, and \cite{Li} for one dimensional surface waves over flat bottoms); this means that the difference between both solutions remains of order $O(\mu^2)$  as long as the wave does not exhibit any kind of singularity such as wave breaking. Near the breaking point, the relevance of the S-GN equations is a completely open problem since the approximations made to derive the non-hydrostatic and dispersive terms (\ref{eqD}) may diverge. Comparison of numerical simulations with experimental data are therefore necessary to assess the validity of the S-GN equations near wave breaking.

The nonlinear shallow water equations (NSWE), or Saint Venant equations, are obtained when the dispersive term $\mu {\bf\cal D}$ is neglected in the S-GN equations. It is well known that these nonlinear hyperbolic equations result in discontinuous solutions (shocks), which can be considered as the mathematical counterparts of breaking wave front. Based on this idea, Hibbert and Peregrine \cite{hp1979} numerically simulated the entire process of bore propagation and runup on a constant beach slope. Kobayashi et al. \cite{koba1989} applied the same approach to simulate the propagation of periodic broken-waves in the surf zone and found good results in comparison with laboratory data. A detailed analysis of the ability of the 1D NSWE shock-wave model to predict cross-shore wave transformation and energy dissipation  in the inner surf  zone was presented by Bonneton \cite{bonn2007}. For 2D problems, Peregrine \cite{peregrine1998} showed that non-uniformities along the breaking wave front (i.e. along the shock) due to alongshore inhomogeneities in the incident wave field or in the local bathymetry, drive vertical vorticity. B\"uhler \cite{buhler2000}  presented a general theoretical  analysis of wave-driven currents and vortex dynamics due to dissipating waves. From computations and laboratory measurements, Brocchini et al. \cite{brocchini2004} and  Kennedy et al \cite{ken2006} showed that breaking-wave-generated vortices are qualitatively well described by the NSWE shock-wave theory. Bonneton et al.  \cite{bonn_DCDS_2009}  emphasized the importance of alongshore inhomogeneities of breaking wave energy dissipation for wave-induced rip current circulation. Their analysis was based on the derivation of an equation for the mean-current vorticity, where the main driving term is related to shock-wave energy dissipation.

The description of shallow water wave dynamics in realistic situations, i.e. over uneven 
bathymetries from the shoaling zone up to the shoreline, requires the development of advanced numerical 
approaches to integrate fully nonlinear Boussinesq-type equations. 
In this framework, the S-GN equations offers the quite exceptional property of admiting closed form solutions of
solitary and cnoidal type, bringing the opportunity to assess the accuracy and
efficiency of numerical methods.

 For horizontal bottoms, the 1D S-GN equations have an exact solitary-wave solution given by, in dimensional variables,
 \begin{subequations}
\begin{equation}
h(x,t)=h_0+H\hspace{0.05cm}\mbox{sech}^2(\kappa(x-ct)),
\end{equation}
\begin{equation}
u(x,t)=c\Big{(}1-\frac{h_0}{h(x,t)}\Big{)},
\end{equation}
\begin{equation}
\kappa=\frac{\sqrt{3H}}{2h_0\sqrt{h_0+H}},
\end{equation}
\begin{equation}
c=\sqrt{g(h+H)},
\end{equation}
\label{sechsoln}
\end{subequations}
where $h_0$ denotes the mean water depth and $H$ the wave height. This family of solutions is known as the Rayleigh solitary wave solution \cite{ray1876}.
Guizien and Barthélémy \cite{guiz2002} experimentally checked that solitary waves generated according to Rayleigh's law display very little dispersive trailing waves compared to KdV ones for instance. 

El et al. \cite{el2006} and  Carter and Cienfuegos \cite{carter2010} have recently shown that the S-GN equations admit also the following family of periodic solutions,
\begin{subequations}
\begin{equation}
h(x,t)=a_0+a_1\mbox{dn}^2\big{(}\kappa (x-ct),k\big{)},
\end{equation}
\begin{equation}
u(x,t)=c\Big{(}1-\frac{h_0}{h(x,t)}\Big{)},
\end{equation}
\begin{equation}
\kappa=\frac{\sqrt{3a_1}}{2\sqrt{a_0(a_0+a_1)(a_0+(1-k^2)a_1)}},
\label{kappa}
\end{equation}
\begin{equation}
c=\frac{\sqrt{ga_0(a_0+a_1)(a_0+(1-k^2)a_1)}}{h_0},
\label{c}
\end{equation}
\label{dnsoln}
\end{subequations}
where $k\in[0,1]$, $a_0>0$, $a_1>0$ are real parameters. 
In equation (\ref{dnsoln}) $\mbox{dn}(\cdot,k)$ is a Jacobi elliptic function with elliptic
modulus $k$. This family of solutions constitutes an important 
extension of the classic KDV cnoidal theory to strongly nonlinear and weakly dispersive applications. 
It is useful to relate the parameters of this solution 
to physical variables in order to compute cnoidal waves in terms of wave
height, $H$, wave period, $T$, and mean water depth, $h_0$. The latter is achieved by solving 
the following system of equations \cite{carter2010},  
%The
%dispersion relation provides a link between celerity and spatial and
%temporal wave scales
%\begin{equation}
%c=\frac{\hat{\omega}}{\hat{k}},
%\label{disp}
%\end{equation}
%where $\hat{\omega}=2\pi/T$ is the angular frequency and $\hat{k}=2\pi/\lambda$ is the wave number.
%Using equations \eqref{dnsoln}-\eqref{eta0} and previous definitions, it is thus possible to find the Serre's 
%cnoidal solution for given $H$, $T$ and $h_0$ by solving
%the following system of equations,
\begin{subequations}
\begin{equation}
a_1=\frac{H}{k^2}
\end{equation}
\begin{equation}
a_0=h_0-a_1 \frac{E(k)}{K(k)}
\end{equation}
\begin{equation}
\hat{\omega}^2=\frac{3\pi^2 g a_1}{4\left[a_0 K(k) + a_1 E(k) \right]^2}
\end{equation}\end{subequations}
where $\hat{\omega}=2\pi/T$ is the angular frequency, 
$K(k)$ and $E(k)$ are the complete elliptic integrals of the first and second kinds respectively.  

As $k\rightarrow 1^-$, the family of periodic solutions limits to the
 two-parameter family of solitary-wave solutions given by equations (\ref{sechsoln}a-d).

%
%%%%%%%%%%%%%%%%%%%%%%%%%%%%%%%%%%%%%%%%%%%%%%%%%%%%%%%%%%%%%%%%%%%%%%%%%%%%%%%%%%%%%%%%%%%%%%%%%%%%%%%%%%%%%%%%%%%%%%%%%%%
%
\section{Reformulations of S-GN equations for numerical implementations}

The  formulation of the S-GN equations in function of conventional unknowns $(h,\bv)$, system (\ref{eqGN}), is not suitable for finite volume methods. 
In this section, we present two other formulations which are convenient for these numerical methods. 

%
%%%
%
\subsection*{S-GN equations in a quasi-conservative form}

In the 1D case, it is possible to show that continuity and momentum equations can be recast in a 
weak
quasi-conservative form by defining an auxiliary
variable $q$ which aggregates all time derivatives in the momentum equations of system (\ref{eqGN}) 
(Cienfuegos et al. \cite{cien_bb_ijnmf_1_2006}). This convenient
mathematical form writes down as,
\begin{eqnarray}
h_t+ F_x&=&0, \nonumber \\
q_t+ G_x &=& \alpha' \mu S',
\label{eqGNpc}
\end{eqnarray}
where the source term in the right hand side is related to the small dispersive 
correction term presented in section \ref{theory},
\begin{equation}
S'=-2 \eps (b-1) b_x \partial_x \left\lbrace uu_x + \zeta_x\right\rbrace.
\end{equation} 

The auxiliary variable reads,
\begin{equation}
q = \left\lbrace 1 + \mu \left[ h_xb_x + b_x^2 + \frac{1}{2} hb_{xx} - hh_x \partial_x
- \left( \frac{h^2}{3} + \alpha' (b-1)^2 \right) \partial_{xx} \right] \right\rbrace u
\label{ufromq}
\end{equation}
and the functions $F$ and $G$ are defined as follows,
\begin{eqnarray*}
F&=&\eps hu,\\
G &=& \eps ( qu + \zeta - \frac{1}{2} u^2) \\ &\ &-\mu \eps \left( \frac{1}{2} b_x^2 u^2 - hb_x uu_x + 
\frac{1}{2} h^2 u^2_{x} 
+ \alpha' (b-1)^2 \lbrack  u^{2}_{x}+  uu_{xx} + \zeta_{xx}\rbrack \right)
\end{eqnarray*}
It is important to note that if no dispersion correction is considered (i.e. $\alpha'=0$) S-GN 
equations can be 
written in the form of conservations law even if bottom variations are allowed.

In the framework of numerical modelling, the system (\ref{eqGNpc}) can be conveniently integrated
over control volumes.

%
%%%
%
\subsection*{S-GN equations in terms of the $(h,h\bv)$ variables} \label{sGNhhv}
An alternative approach proposed by \cite{bclmt2009} is to write the S-GN equations in terms of the conservative variables $(h,h\bv)$, namely 
\begin{eqnarray} \nonumber
h_t+\eps \nabla\cdot(h\bv)&=&0\nonumber\\
(h\bv)_t+\eps \nabla\cdot(h\bv \otimes\bv) + h\nabla\zeta &=& -(I+\mu \alpha h{\mathcal T}[h,b] \frac{1}{h})^{-1}\big[ \frac{1}{\alpha} h\nabla\zeta \nonumber\\ 
& & \quad + \eps\mu h \cQ_1[h,b](\bv) \big] + \frac{1}{\alpha}h\nabla\zeta
\label{eqGNhhv}
\end{eqnarray}
with $\cQ_1[h,b](\bv)=\cQ[h,b](\bv)-{\mathcal T}[h,b]((\bv\cdot\nabla)\bv)$. It is worth pointing out that this formulation does not include any third-order derivative, allowing for easier and robust numerical computations, especially when the wave becomes steeper. Note that the S-GN equations with improved dispersion \`a la Nwogu can also be put under a similar form \cite{CLM}.

This formulation is well-suited for a splitting approach with a finite volume method for the hyperbolic part of the equations (the left-hand side of  system (\ref{eqGNhhv}))  and finite difference method for the dispersive part (the right-hand side of  system (\ref{eqGNhhv})).
%
%%%%%%%%%%%%%%%%%%%%%%%%%%%%%%%%%%%%%%%%%%%%%%%%%%%%%%%%%%%%%%%%%%%%%%%%%%%%%%%%%%%%%%%%%%%%%%%%%%%%%%%%%%%%%%%%%%%%%%%%%%%
%

%
%%%%%%%%%%%%%%%%%%%%%%%%%%%%%%%%%%%%%%%%%%%%%%%%%%%%%%%%%%%%%%%%%%%%%%%%%%%%%%%%%%%%%%%%%%%%%%%%%%%%%%%%%%%%%%%%%%%%%%%%%%%
%
\section{High-order compact finite volume method}

In the 1D case, the quasi-conservative form \eqref{eqGNpc} can be numerically integrated to 
describe wave propagation in shallow waters. However, in order to 
extend the application of the model into the surf zone, additional terms have to be added to the 
mass and momentum conservation equations. These terms aim at modelling wave-breaking energy dissipation 
and bottom friction. In dimensional variables, this extended system can be written in the following form,
% ----------------------------------------------------------------------------
\begin{eqnarray}
h_t+ F_x&=& D_h, \nonumber
\label{eqmass}\\
q_t+ G_x &=& \frac{1}{h} D_{hu} - \frac{\tau_b}{\rho h} + \alpha' S',
\label{eqqdm}
\end{eqnarray}
% ----------------------------------------------------------------------------
where $\rho$ is the water density, $D_h$ and $D_{hu}$ represent breaking terms, 
$\tau_b$ is the bed shear stress.   

Breaking-induced energy dissipation mechanisms are introduced through diffusive-like terms, $D_h$ and $D_{hu}$,
applied locally on the wave front face where an explicit breaking criterion is required to switch them on. 
The mathematical form for $D_h$ and $D_{hu}$ is chosen in order to ensure that the overall mass and momentum budget 
is preserved, acting only as to locally redistribute these quantities under the breaker \cite{cien2009}. 
Breaking terms are thus written in the form,
% ------------------------------------------------------------------------------
\begin{gather*}
D_{h}= \partial_x\left( \nu_h h_x \right),
\label{eqDiss1} \\
D_{hu}= \partial_x \left( \nu_{hu} (hu)_x \right),
\label{eqDiss2}
\end{gather*} 
% ------------------------------------------------------------------------------
where $\nu_h$ and $\nu_{hu}$ are diffusivity functions expressed as, 
% ------------------------------------------------------------------------------
\begin{gather*}
\nu_h (X) = - K_h \exp \left ( \frac{X}{l_r} -1 \right ) \left \lbrack \left (\frac{X}{l_r} -1 \right) + \left (\frac{X}{l_r} -1 \right )^2 \right \rbrack,
\label{nu1}\\                        
\nu_{hu} (X) = - K_{hu} \exp \left ( \frac{X}{l_r} -1 \right ) \left \lbrack \left (\frac{X}{l_r} -1 \right) + \left (\frac{X}{l_r} -1 \right )^2 \right \rbrack,
\label{nu2}                        
\end{gather*}   
% ------------------------------------------------------------------------------
with $K_h$ and $K_{hu}$ slowing varying scaling coefficients, $X$ is a moving horizontal 
coordinate attached to the wave crest
and $l_r$ is the extent over which breaking terms are active. This breaking model has been calibrated on 
the Ting and Kirby's \cite{ting94} regular wave experiment and optimal parameter values suggested 
by Cienfuegos et al. \cite{cien2009} are $K_h=2cd$, 
$K_{hu}=20cd$ and $l_r/d=0.82$, with $c=(gd)^{0.5}$ and $d$ the local still water depth.

The numerical integration of the system \eqref{eqqdm} is performed using a 
high-order compact finite volume method. A detailed description of this model, SERR-1D, 
is given in references \cite{cien_bb_ijnmf_1_2006,cien_bb_ijnmf_2_2007}. 
The equations are first integrated in space over discrete control volumes 
$\Omega_{i}=\lbrace x \in [x_{i-\frac{1}{2}},x_{i+\frac{1}{2}}] \rbrace$, 
% -------------------------------------------------------------------------------------------
\begin{gather}
\frac{\partial}{\partial t}\int_{x_{i-\frac{1}{2}}}^{x_{i+\frac{1}{2}}} h\,dx + F(x_{i+\frac{1}{2}},t) 
- F(x_{i-\frac{1}{2}},t)= \int_{x_{i-\frac{1}{2}}}^{x_{i+\frac{1}{2}}} \partial_x\left( \nu_h h_x \right)\,dx,
\label{eq:intcont}\\
\frac{\partial}{\partial t}\int_{x_{i-\frac{1}{2}}}^{x_{i+\frac{1}{2}}} q\,dx + G(x_{i+\frac{1}{2}},t) - 
G(x_{i-\frac{1}{2}},t)=\int_{x_{i-\frac{1}{2}}}^{x_{i+\frac{1}{2}}} \left(S'+
\frac{1}{h}\partial_x \left( \nu_{hu} (hu)_x \right)-\frac{\tau_b}{\rho h}\right)\,dx,
\label{eq:intmom}
\end{gather}
% -------------------------------------------------------------------------------------------
where integral of variables $h$ and $q$ over control volumes must be advanced in time. 
The average value of function $h$ at time $t=t_n$ 
over control volume $\Omega_i$ is noted as,
% -------------------------------------------------------------------------------------------
\begin{gather*}
\widehat{h}^n_{i}=\frac{1}{\Delta x}\int_{x_{i-\frac{1}{2}}}^{x_{i+\frac{1}{2}}} h(x,t_n)\,dx.
\label{eq:cellaver}
\end{gather*}
% -------------------------------------------------------------------------------------------
where $\Delta x=x_{i+\frac{1}{2}}-x_{i-\frac{1}{2}}$ is the length of the discrete control volumes.
Hence, integrated over the whole domain, equations \eqref{eq:intcont} and \eqref{eq:intmom} 
can be expressed as,
% -------------------------------------------------------------------------------------------
\begin{eqnarray*}
\frac{d\, \widehat{h}^n_{i}}{dt} &= &- \frac{1}{\Delta x} \left ( F^n_{i+\frac{1}{2}} - F^n_{i-\frac{1}{2}} 
- \lbrace \nu_h h_x\rbrace |_{i+\frac{1}{2}} + \lbrace \nu_h h_x\rbrace |_{i-\frac{1}{2}} \right ),
 \ \ \ \ 
\label{eq:intcont2}\\
\frac{d\, \widehat{q}^n_{i}}{d t} &=&  \widehat{S}^n_{i} - \frac{1}{\Delta x} \left ( G^n_{i+\frac{1}{2}} - 
G^n_{i-\frac{1}{2}} \right ), \ \ \ \ \ \ \ \ \ \ \  \text{for}\ \ i=1,2,...,N,
\label{eq:intmom2}
\end{eqnarray*}
% -------------------------------------------------------------------------------------------
where $N$ is the total number of control volumes used to discretize the physical domain and
$\widehat{S}^n_{i}$ is the discretized counterpart of the source term in the right hand side 
of equation \eqref{eq:intmom}. This term is approximated through 
centred finite differences. The values $h$ and $q$ at cell interfaces  are reconstructed from
cell-averaged values using the implicit 4th order compact interpolation technique described in 
references \cite{koba99,laco04}. At each time step, the velocity component at control volume interfaces 
is computed numerically by inverting equation \eqref{ufromq}. Time stepping is performed through 
a 4th order Runge-Kutta method.

Efficient absorbing-generating boundary conditions have been implemented in SERR-1D. They are based on
the following characteristic form of the S-GN equations,
% -------------------------------------------------------------------------------------------
\begin{gather}
\frac{d R^{+}}{dt} = - \frac{1}{3h} \frac{\partial}{\partial x} \left ( h^2 P\right ) 
- gb_x - \frac{\tau_b}{\rho h} \quad \text{along} \quad \frac{dx}{dt} = u + \sqrt{gh},
\label{eq:qcarac1}\\
\frac{d R^{-}}{dt} = - \frac{1}{3h} \frac{\partial}{\partial x} \left ( h^2 P\right ) 
- gb_x - \frac{\tau_b}{\rho h} \quad \text{along} \quad \frac{dx}{dt} = u - \sqrt{gh},
\label{eq:qcarac2}
\end{gather}
% -------------------------------------------------------------------------------------------
with positive and negative Riemann variables defined respectively as $R^{+} = u + 2 \sqrt{gh}$ and 
$R^{-} = u - 2 \sqrt{gh}$. Vertical acceleration of fluid particles, 
which is of order $O(\mu)$ and thus disregarded in the nonlinear shallow water equations (NSWE), is represented by 
function $P$ in the first term of the right hand side of equations \eqref{eq:qcarac1}-\eqref{eq:qcarac2}. This
term is responsible for the loss of hyperbolicity in the S-GN equations by 
introducing an horizontal dependence in the characteristic plane $(x,t)$. 

It is worth noting that from a physical point of view we can expect that time scales associated 
to dispersive effects would be larger than the ones associated to nonlinearities from intermediate to shallow 
waters. 
We may therefore assume that over short distances/times, Riemann variables 
might be locally conserved along characteristics. This physical argument 
has been used to develop absorbing-generating boundary conditions for the numerical  
resolution of S-GN  equations \cite{cien_bb_ijnmf_2_2007,mignot2009}. 

For the moving shoreline boundary condition, the simple extrapolation technique proposed by 
Lynett et al. \cite{lyne02} has been adapted in the finite volume resolution.

SERR-1D has been extensively validated by comparisons with non-breaking and breaking wave laboratory experiments \cite{cien_ICCE2006,cien_bb_ijnmf_2_2007,cien2009,mignot2009}.

\begin{figure}[h]
\centering\includegraphics[width=13cm,keepaspectratio=true]{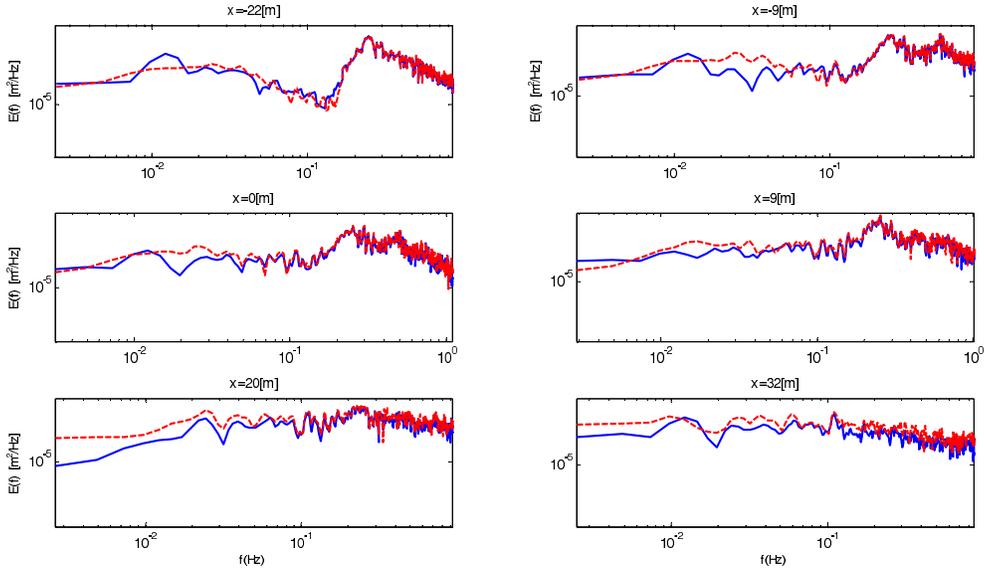}
\caption{Random wave transformation and breaking over a mild slope beach. The origin 
of the x-coordinate is at the toe of the beach slope, positive towards the shore. 
Blue : experimental data. Red : computed results.}\label{validation2}
\end{figure}

In the present paper, the capabilities of the model are illustrated by comparing numerical computations 
with breaking random  wave propagation experiments. We use measurements conducted at the 70-meter long wave tank 
of the Instituto Nacional de Hidraulica (Chile), prepared with a beach of very mild slope of 1/80 
in order to produce large surf zone extensions \cite{duarte2008}. A random JONSWAP type wave field (h0=0.52m, 
fp=0.25Hz, Hmo=0.17m)
was generated by a piston wave-maker and measurements of the free surface displacements were performed all over its
length at high spatial resolution (0.2m to 1m). This experiment allows us to test numerical models
that are assumed to reproduce nonlinear shallow water wave propagation, breaking and run-up.
The evolution of the wave energy power spectral density as the wave field propagates over the beach
is presented in Figure \ref{validation2}, for both experimental measurements and numerical results. 
SERR-1D is able to simulate the complex nonlinear energy transfer occuring in the shoaling and surf zones. 
In particular, it reproduces the generation of higher frequency harmonics during shoaling (between x=-9m and x=9m), 
energy dissipation by breaking in the surf zone (between x=9m and x=32m). The numerical results also
indicate that the model is able to reproduce the energy transfer from the Jonswap spectrum band ($>$0.1Hz)
to the infragravity band ($<$0.1Hz). It is important to note that the numerical model was forced with the 
high-pass filtered wave signal measured two meter away from the wave paddle without energy content in 
the infragravity band.

%
%%%%%%%%%%%%%%%%%%%%%%%%%%%%%%%%%%%%%%%%%%%%%%%%%%%%%%%%%%%%%%%%%%%%%%%%%%%%%%%%%%%%%%%%%%%%%%%%%%%%%%%%%%%%%%%%%%%%%%%%%%%
%
\section{High-order hybrid finite volume / finite difference method}
The formulation of S-GN equations introduced in section \ref{sGNhhv} (equations (\ref{eqGNhhv})) is well-suited for a splitting approach separating the hyperbolic and the dispersive part of the equations. In this section, we first present an efficient high-order positive preserving well-balanced shock-capturing scheme for the hyperbolic step and then the splitting method for solving the whole S-GN system.  

%%%%%%%%%%%%%%%%%%%%%%%%%%%%%%%%%%%%%%%%%%%%%
%
\subsection*{NSWE shock capturing solver}

We consider in this section the hyperbolic part of the S-GN equation, stated in the dimensionnalized form:
\begin{eqnarray}\label{eq_sv1}
h_t+\nabla\cdot(h\bv)&=&0 \\
(h\bv)_t+ \nabla\cdot(h\bv \otimes\bv)+gh\nabla\zeta &=& 0.
\nonumber
\end{eqnarray} 
This system can be also regarded as an hyperbolic system of conservation laws with a geometrical source term controlled by the topography variations. To simplify the algorithm presentation we only consider the one-dimensional problem:
\begin{equation}\label{sv1d}
\vw_t+ f(\vw)_x=S(\vw),
\end{equation}
where $\vw = {}^t\left(h, hu\right)$, $f(\vw) = {}^t\left(hu, hu^2+0.5gh^2\right)$ and $S = {}^t\left(0, -ghb_x\right)$ the source term.
To perform numerical approximations of the weak solutions of this system, we use a high order finite-volume approach in conservative variables, relying on Riemann problems for hyperbolic conservative laws \cite{godlewski_raviart}. This approach allows accurate computation of propagating bores, with reduced spurious effects of numerical dissipation and dispersion. Using such accurate scheme, we are able to handle wave breaking (see Section \ref{theory}). Since we aim at computing the complex interactions between propagating waves and topography (including the preservation of motionless steady states), we also embed this approach into a well-balanced scheme.

More precisely, based on discrete finite-volume cell averaging $\bar{\vw}_i^n$ at time $t^n=n\delta t$, we use the limited $4^{th}$-order  MUSCL reconstruction suggested in \cite{bert2008}. Considering a cell $C_i$, this approach provides, for all $t^n$, high order accuracy interpolated quantities $\bar{\vw}_{i,l}$ and $\bar{\vw}_{i,r}$, respectively at the left and right boundary of each cell.
To get a positive preserving and well-balanced scheme, additional faces reconstructions are introduced \cite{audusse}:
$$
\left. \begin{array}{lll}
 b_i^*  & =  &\max(b_{i,r}, b_{i+1,l}),\\
h_{i,r}^* & =  &\max(0, h_{i,r} + b_{i,r} - b_i^*),\\
h_{i+1,l}^* &= &\max(0, h_{i+1,l} + b_{i+1,l} - b_i^*).
\end{array}\right.
$$
These new left and right values for water height are used to compute auxiliary conservative faces values $\vw_{i,r}^*$ and $\vw_{i+1,l}^*$:
\be
\vw_{i,r}^* = \left(\begin{array}{c}
h_{i,r}^*\\
h_{i,r}^* u_{i,r}
\end{array}\right),
\quad
\vw_{i+1,l}^* =\left( \begin{array}{c}
h_{i+1,l}^*\\
h_{i+1,l}^* u_{i+1,l}
\end{array}\right)
\ee
which are  injected into a Riemann solver. Neglecting temporally the time discretization, we obtain the following semi-discrete finite-volume scheme for (\ref{sv1d}):
$$
\frac{d}{dt}\bar{\vw}_i (t) +   \frac{1}{\Delta x}   \Bigl( \vf_{i+\fr 1 2}  \bigl(\bar{\vw}_{i,r},  \bar{\vw}_{i+1,l}, b_{i,r}, b_{i+1,l}
                    \bigr)   -  \vf_{i-\fr 1 2} \left(\bar{\vw}_{i-1,r},\bar{\vw}_{i,l}, b_{i-1,r}, b_{i,l}  \right) \Bigr)=  S_{c,i}
$$
where $\vf_{i+\fr 1 2}$ and $\vf_{i-\fr 1 2}$ are the numerical flux functions based both on a conservative flux consistent with the homogeneous NSWE issued from a relaxation approach \cite{bert2008}, and the hydrostatic reconstruction correction to the interface fluxes \cite{audusse}.  $S_{c,i}$ is a centered discretization of the source term needed to achieve  accuracy, consistency and well-balancing properties. \\

The resulting finite-volume scheme provides high-order accuracy approximations of the weak solutions of system (\ref{sv1d}) while preserving the positivity of the water-height, thanks to the relaxation approach developed in \cite{bert2008}. This last property is paramount to ensure robust and accurate simulations of time-evolving shorelines \cite{marc2007}. In addition, the use of a well-balanced scheme leads to accurate computations of incident waves and topography complex interactions classically occurring in the surf zone \cite{marc_ICCE2006}. It also allows far more accurate results in situations involving tiny oscillations near steady states. The resulting high-order positive preserving well-balanced shock-capturing scheme was implemented in the code SURF-WB \cite{marc2007,bert2008}.

\begin{figure}[h]
\includegraphics[width=6cm,keepaspectratio=true]{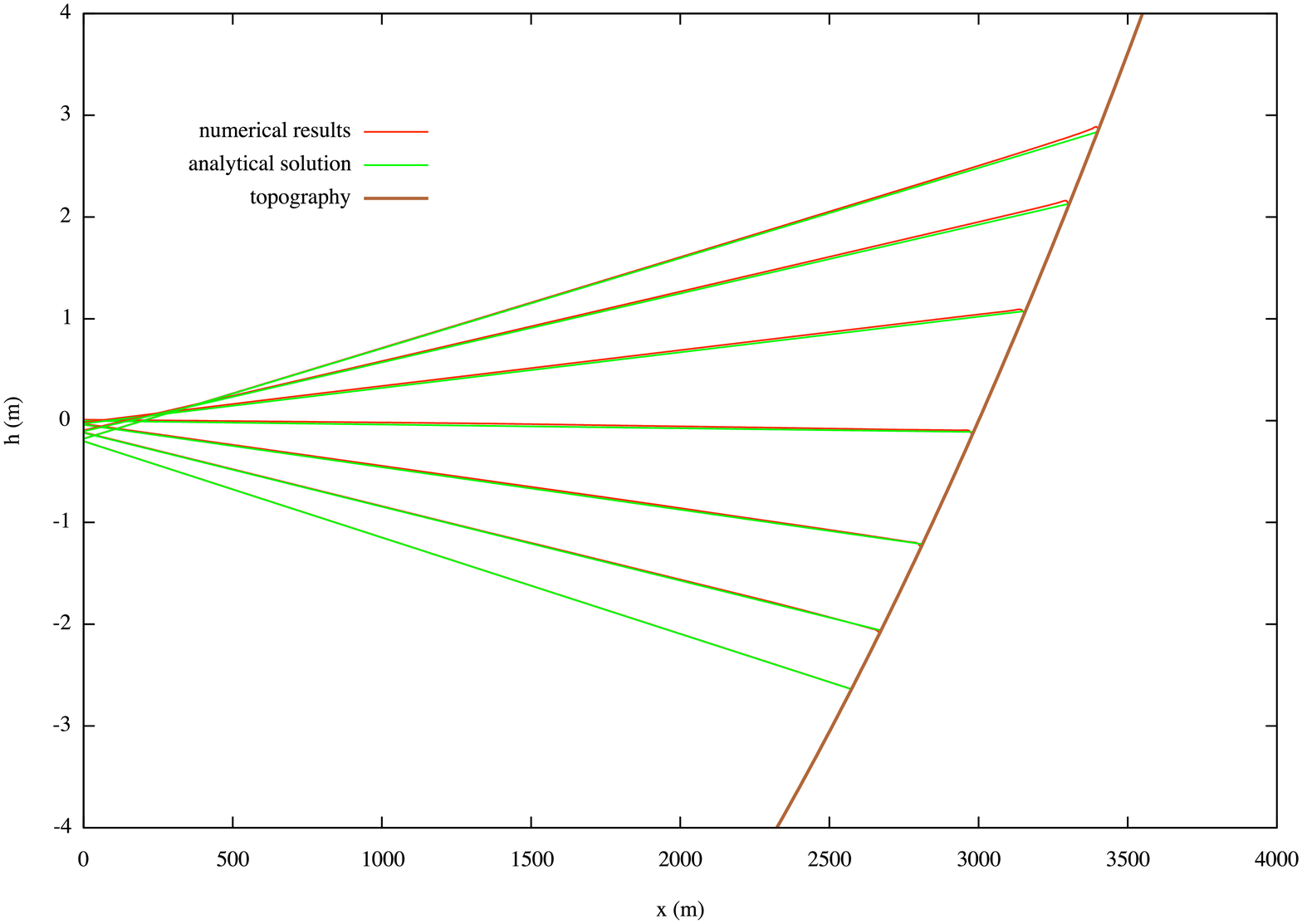}
\includegraphics[width=6cm,keepaspectratio=true]{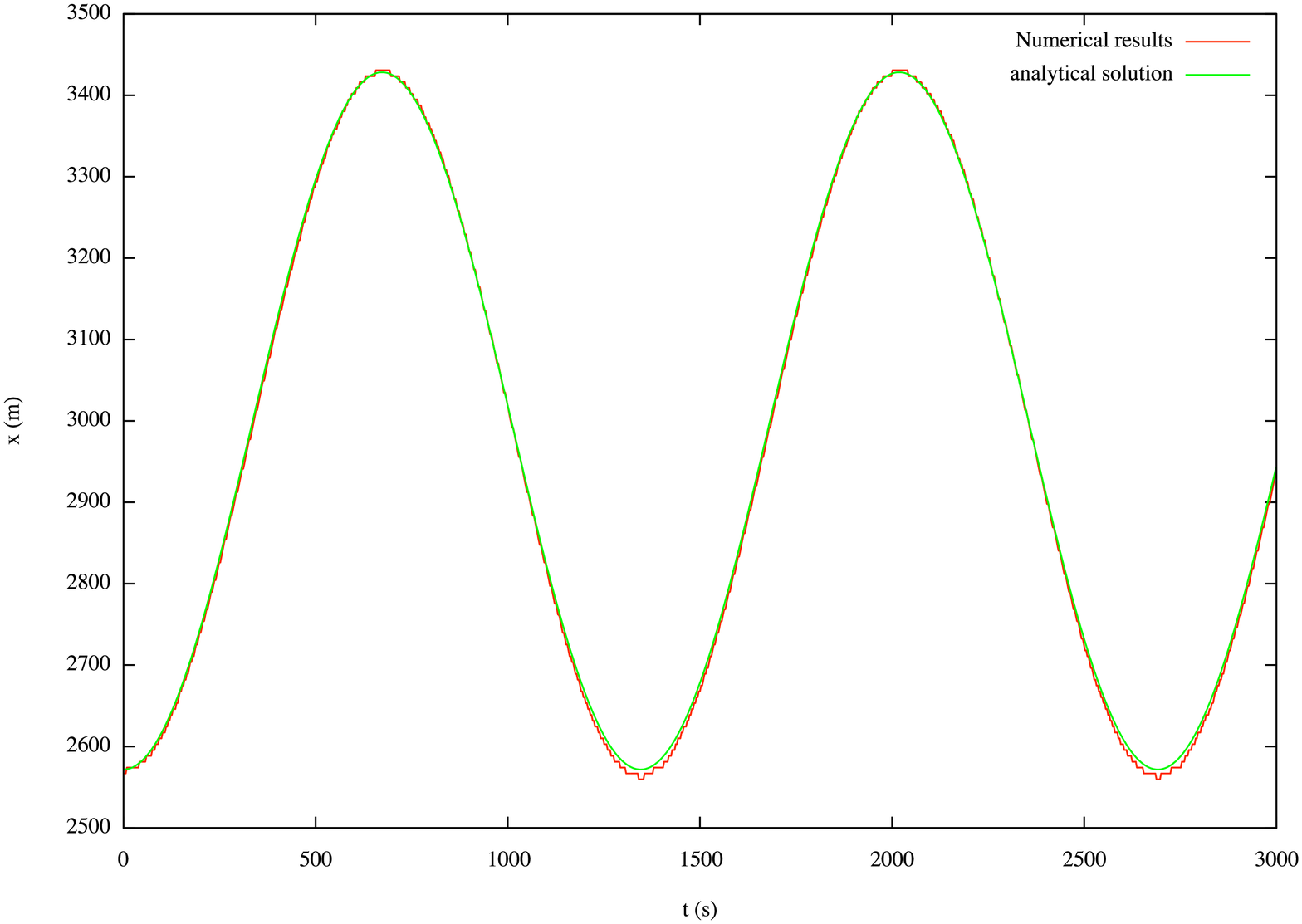}
\caption{(a) Free surface at various time between a half evolution period (analytical solution in solid line) (b) Shoreline location between $t=0\,s$ and $t=3000\,s$}\label{validation3}
\end{figure}

We can observe on Figure \ref{validation3} a comparison between numerical results and analytical solutions for a one dimensional test case. This solution describes time oscillations of a forced flow over a quadratic topography profile, involving a moving shoreline \cite{sampson2007}. The free surface is always planar during the oscillations. The computational domain is $4320\,m$ long and the topography is given by $$
b(x) = 1-h_0\left(   \frac{x}{a}\right)^2
$$
where $h_0=10\,m$ and $a=3000\,m$ are respectively the mean water depth and a topography scaling parameter.
The flow motion is driven by the left boundary condition, where we impose a periodic motion:
\begin{equation}
h(t,0) = -\frac{a^2 B^2}{8 g^2 h_0}  \left(\frac{8gh_0}{a^2}\cos(2\sqrt{8gh_0/a^2}t)   \right) - \frac{B^2}{4 g} 
\end{equation}
where $B$ is a free parameter, set to $2\,m.s^{-1}$ for the simulation shown here.
Note that the shoreline location, $x_s$, can be analytically derived:
$$x_s = \frac{a^2 }{2gh_0}\left( -B\sqrt{8gh_0/a^2} \cos(\sqrt{8gh_0/a^2}t)  \right) + a
$$
The simulation has been performed with $200$ cells and the time step is set to $0.03\,s$, with a first order scheme. Results are shown first as a comparison between numerical results and analytical solution for the free surface elevation, at several times. Note that the results and the solution are almost indistinguishable. Then we highlight the accuracy of the shoreline location prediction through a comparison between analytical and numerically predicted shoreline location, during $3000\,s$.

%
%%%%%%%%%%%%%%%%%%%%%%%%%%%%%%%%%%%%%%%%%%%%%
%
\subsection*{S-GN splitting solver}

Equations (\ref{eqGNhhv}) are solved using the following splitting method : we decompose the solution operator $S(\cdot)$ associated to (\ref{eqGNhhv}) at each time step by the second order splitting scheme
\begin{equation}\label{S0}
	S(\delta_t)=S_1(\delta_t/2)S_2(\delta_t)S_1(\delta_t/2),
\end{equation}
where $S_1(t)$ is the solution operator associated to the nonlinear 
shallow water equations (\ref{eq_sv1}), while $S_2(t)$ is the solution operator associated to 
the dispersive part of the equations, namely
\be
\label{S2}
\left\lbrace
\begin{array}{lcl}
h_t &=&0,\\
(h V)_t &=& - (I+\alpha h\cT\frac{1}{h})^{-1}\big[\frac{1}{\alpha} gh \nabla\zeta + h\cQ_1(V)\big] + \frac{1}{\alpha} g h \nabla \zeta.
\end{array}\right.
\ee

As described previously, $S_1(t)$ is computed using a finite-volume approach, while a finite-difference approach is used to solve $S_2(t)$ at each time step: the spatial derivatives are discretized using centered fourth-order formulae. Boundary conditions are imposed using the method presented in \cite{bclmt2009}. As far as time discretization is concerned, we choose to solve $S_1(t)$ and $S_2(t)$ using an explicit fourth-order Runge-Kutta scheme.\\
To sum up, $S_1(t)$ and $S_2(t)$ are solved, within our splitting approach, using a fourth-order scheme in space and time. However, the use of a second-order splitting method implies that the global scheme is of order two in time.  The use of a fourth-order Runge-Kutta scheme for $S_1(t)$ and $S_2(t)$  is however required to have a good semi-discrete dispersion relation (see \cite{bclmt2009}).

In order to handle wave breaking, we switch from the S-GN equations to the NSWE, locally in time and space, by skipping the dispersive step $S_2(\delta t)$ when the wave is ready to break. In this way, we only solve the hyperbolic part of the equations for the wave fronts, and the breaking wave dissipation is represented by the shock energy dissipation (see also  \cite{bonn2007}).  
To determine where to suppress the dispersive step at each time step, we use the first half-time step $S_1$ of the time-splitting as a predictor to assess the local energy dissipation.
 This dissipation is close to zero in regular wave regions, and forms a peak when shocks are appearing. We can then easily locate the eventual breaking wave fronts at each time step, and skip the dispersive step only at the wave fronts. \\

%%%%%%%%%%%% PROPAGATION OF A STRONGLY NON LINEAR CNOIDAL WAVE %%%%%%%%%%%%%%%%%%%%%%%%%
In order to test the efficiency of our numerical methods, the ability of the model to describe the propagation of a strongly non-linear cnoidal wave solution of the S-GN equations (cf. relations (\ref{dnsoln})) is investigated. Periodic boundary conditions have been used in order to appreciate the propagation of the cnoidal wave at long time, and the computational domain length is equal to the cnoidal wave-length.

\begin{figure}[h]
\centering
\includegraphics*[width=8cm]{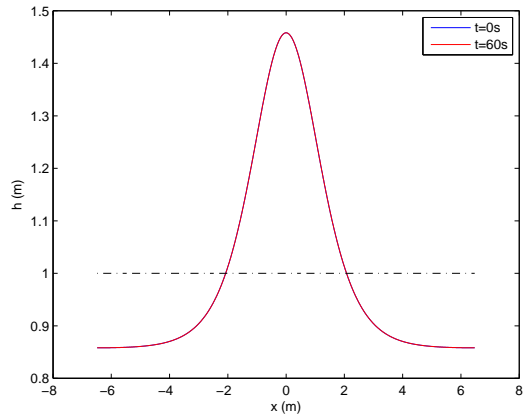}
\caption{Propagation of a strongly nonlinear cnoidal wave over a periodic domain. At $t=0\, s$ (blue line), the cnoidal wave is such as $H = 0.6 \,m$, $h_0 = 1\,m$ and $T=4 \, s$.  The red line represents the wave 15 periods later ($t=60 \, s$). The doted line is the still water level. $dx = 0.01 \,m$ and the Courant number is equal to 1.}\label{validation4}
\end{figure}

The initial condition is a cnoidal wave with $H = 0.6$m, $h_0 = 1$m and $T=4$s (see Figure \ref{validation4}). Numerical and theoretical solutions are compared at $t=15T=60$s.
Relative amplitude and celerity errors are given in Table \ref{tab_erreur} for different spatial steps. The Courant number remains equal to 1 for the different cases.

Figure \ref{validation4} shows the results for the finest grid size considered ($dx = 0.01$m). The cnoidal wave at $t=0$s and $t=60$s are both plotted in this figure, but cannot be distinguished since the errors on wave height and celerity are extremely small (cf. Table \ref{tab_erreur}). In particular,   
the celerity error cannot be precisely quantified for $dx = 0.01$m since the spatial lag between numerical and analytical solution at $t=60$s is much smaller than the grid size : the numerical solution converge to the exact one for very small $dx$. 
Moreover,  relative celerity and amplitude errors remain small for larger spatial steps, demonstrating the very high accuracy of our numerical methods.
\\

\begin{table}[h]
\begin{center}
\begin{tabular}{|c|c|c|}
     \hline
               & Relative Amplitude Error (\%) & Relative Celerity Error (\%) \\
     \hline          
     dx = 0.01 m & 3.1 $10^{-5}$ & $ < 5.4$ $10^{-3}$ \\
     \hline
     dx = 0.10 m & -0.24 & 0.026 \\
     \hline
     dx = 0.15 m & -1.75 &  0.031 \\
     \hline
\end{tabular}
\end{center}
\caption{ Amplitude and  celerity errors relative to the analytical solution for the cnoidal wave previously described (cf. Figure \ref{validation4}) at $t = 60$s. Computations are performed for 3 spatial steps and a Courant number equal to 1. \label{tab_erreur}}
\end{table}

%%%%%%%%%%%%%%%%%%%%%	 SYNOLAKIS	 %%%%%%%%%%%%%%%%
In the next case,  we assess the ability of our model to describe wave runup and breaking. It is based on laboratory experiments carried out by Synolakis  \cite{syno}, for an incident solitary wave of relative amplitude $a_0/h_0 = 0.28$, propagating and breaking over a planar beach with a slope of 1:19.85.
% Free surface elevations at different times are available thanks to video measurements. 
The still water level in the horizontal part of the beach was $h_0=0.3$m, and the simulations are performed using the grid size $dx = 0.08$m and  $dt = 0.02$s. 
As friction effects are important when the water becomes very shallow, in particular for the run-up and run-down stage, a quadratic friction term is introduced for this simulation. 

\begin{figure}
\centering
\includegraphics*[width=13cm]{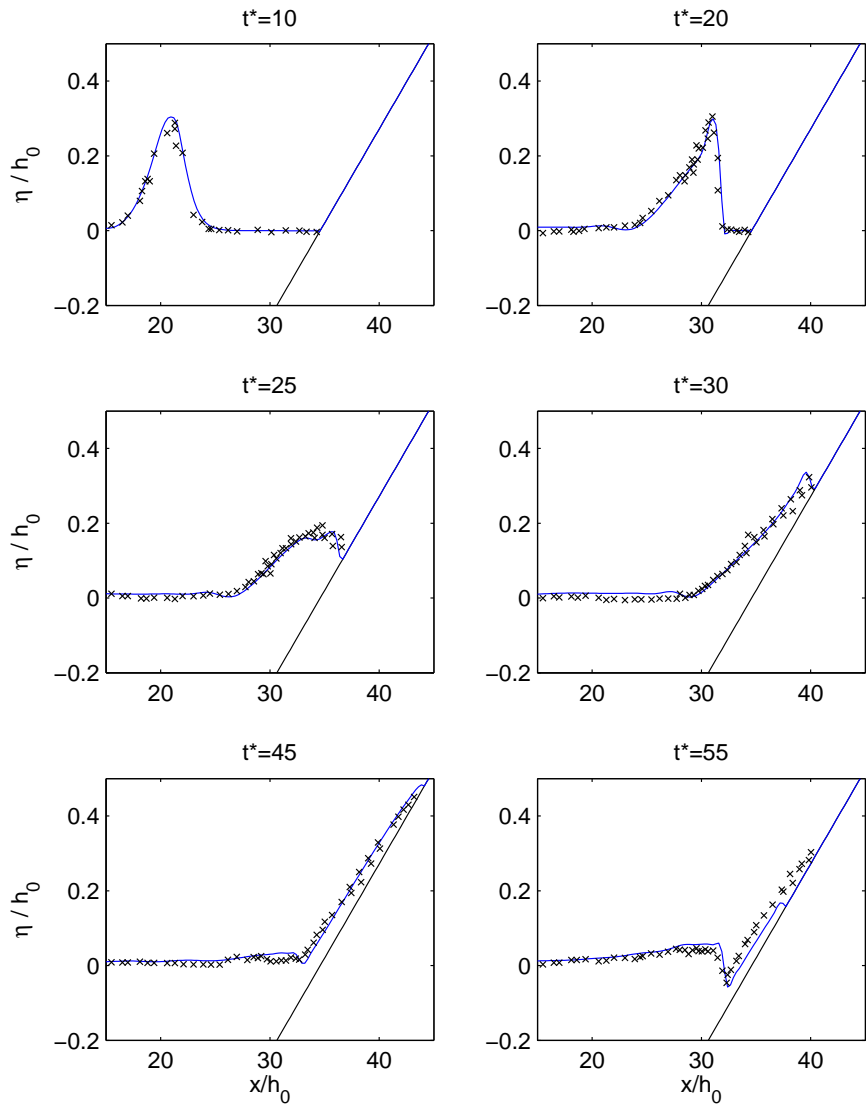}
\caption{Comparisons of model predictions (---) and experimental data (x) for a breaking
solitary wave with non-dimensional initial incident amplitude $ a_0 /h_0 = 0.28$, on a $1:19.85$ constant slope
beach investigated by Synolakis (1987). $t^*=t(g/h_0)^{1/2}$. Adapted from \cite{bclmt2009}. }\label{validation5}
\end{figure}

The comparison between measured and computed waves is presented in Figure \ref{validation5}. It shows a good agreement between model predictions and laboratory data for the wave shoaling, breaking, run-up and run-down. In particular, the model is able to accurately describe the formation and breaking of the back-wash bore without any additional treatment, which is a particularly demanding test.

% and illustrates the ability of our model to reproduce shoaling, breaking, run-up and run-down, as well as the formation and breaking of the backwash bore, and this without any additional treatment.

%

%%%%%%%%%%%%%%%%%%%%%%%%%%%%%%%%%%%%%%%%%%%%%%%%%%%%%%%%%%%%%%%%%%%%%%%%%%%%%%%%%%%%%%%%%%%%%%%%%%%%%%%%%%%%%%%%%%%%%%%%%%%
%
\section{Conclusion}
As the waves approach the shore, the nonlinearity  effects become intense, especially in  the final stages of shoaling and the surf zone. To simulate such nonlinear processes in shallow water, fully nonlinear Boussinesq-type approaches are required. The Serre or Green Naghdi (S-GN) equations, with improved dispersion properties, represent the relevant system to model these highly nonlinear weakly dispersive waves (see Lannes and Bonneton \cite{lann_POF2009}). Two high-order methods for solving S-GN equations, based on Finite Volume approaches, are presented in this article. 

The first one is based on a quasi-conservative form of  the S-GN equations. Wave-breaking energy dissipation is taken into account through a diffusive-type parametrization on both the mass conservation and momentum conservation \cite{cien2009}. This model, SERR-1D, has been extensively validated by multiple comparisons between numerical simulations and physical experiments including solitary waves shoaling, regular waves propagating over a submerged bar \cite{cien_bb_ijnmf_2_2007}  or regular wave breaking over uniform beach slopes \cite{cien2009}. In the present paper,  new results show the ability of the model to reproduce nonlinear energy transfer for random waves, in the shoaling and surf zones.

We present also an alternative approach where the S-GN equations are reformulated in terms of the conservative variables $(h,h\bv)$ (equations (\ref{eqGNhhv})). This formulation is well-suited for a splitting approach with a finite volume method for the hyperbolic part of the S-GN equations (NSWE) and a finite difference method for the dispersive part \cite{bclmt2009}. Our hyperbolic method is based on a high-order well-balanced shock-capturing scheme (SURF-WB code, \cite{marc2007,bert2008}). In the present article we show that our hybrid model accurately describes strongly nonlinear S-GN cnoidal wave solutions.  In order to handle wave breaking, we switch locally from the S-GN equations to the hyperbolic NSWE, where the wave is ready to break. In this way, we only solve the hyperbolic part of the equations for the wave fronts, and the breaking wave dissipation is represented by shock energy dissipation. We show that this approach accurately predicts nonlinear shoaling, breaking and runup of solitary waves on a beach. The advantage of this approach, in comparison with classical breaking parametrizations, is that it is easily extended to 2DH broken-wave problems. This is crucial to predict wave-induced circulations and macro-vortices, which are strongly controlled by dissipation non-uniformities along broken-wave fronts \cite{broc2008,bonn_DCDS_2009}.

\begin{figure} 
\begin{center}
\includegraphics[width=12cm]{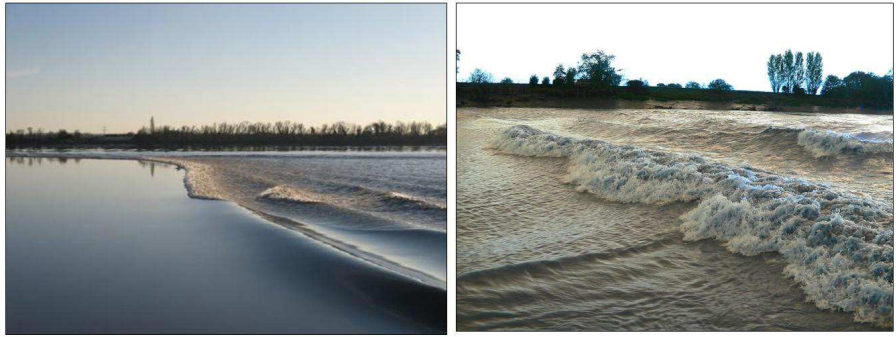}
\caption{Tidal bore propagation in Garonne River; Mascaret project (Parisot and Bonneton, 2009-2011).}
\label{mascaret}
\end{center}
\end{figure}

Further work is required to evaluate the 
capability of our S-GN models to predict such 2DH flows. An another important open problem concerns the ability of our approaches to predict bore dynamics in a large range of Froude numbers, from "non-breaking undular bore" to "breaking  bore". The transition between those two types of bores, which is controlled by the competition between nonlinearity, dispersion and dissipation effects, is still poorly understood. Recent field experiments on the dynamics of  tidal bores (see figure \ref{mascaret}) will give us the opportunity to validate our models.

%
%%%%%%%%%%%%%%%%%%%%%%%%%%%%%%%%%%%%%%%%%%%%%%%%%%%%%%%%%%%%%%%%%%%%%%%%%%%%%%%%%%%%%%%%%%%%%%%%%%%%%%%%%%%%%%%%%%%%%%%%%%%
%
\section*{Acknowledgements} 
The authors would like to acknowledge the financial and scientific support of the French INSU - CNRS (Institut National des Sciences de  l'Univers -  Centre National de la Recherche Scientifique) program IDAO ( "Interactions et Dynamique de l'Atmosphère et de l'Océan ". This work has also been supported by the
ANR MathOcean, the ANR MISEEVA and the project ECOS-CONYCIT action C07U01.


\bibliographystyle{elsarticle-num}



\end{document}